\documentclass[aps,prl,twocolumn,superscriptaddress]{revtex4}
\usepackage{graphicx}
\usepackage{amsmath}
\usepackage{amssymb}
\begin{document}

\bibliographystyle{apsrev}

\title{Tunable Superconducting Phase Transition in Metal-Decorated Graphene Sheets}

\author{B. M. Kessler}
\affiliation{Department of Physics, University of California at Berkeley, Berkeley, CA, 94720  USA.}
\affiliation{Materials Sciences Division, Lawrence Berkeley National Laboratory, Berkeley, CA, 94720  USA.}
\author{\c{C}. \"O. Girit}
\affiliation{Department of Physics, University of California at Berkeley, Berkeley, CA, 94720  USA.}
\affiliation{Materials Sciences Division, Lawrence Berkeley National Laboratory, Berkeley, CA, 94720  USA.}
\author{A. Zettl}
\affiliation{Department of Physics, University of California at Berkeley, Berkeley, CA, 94720  USA.}
\affiliation{Materials Sciences Division, Lawrence Berkeley National Laboratory, Berkeley, CA, 94720  USA.}
\affiliation{Center of Integrated Nanomechanical Systems, Berkeley, CA,94720 USA.}
\author{V. Bouchiat}
\affiliation{Department of Physics, University of California at Berkeley, Berkeley, CA, 94720 USA.}
\affiliation{Institut N\'eel, CNRS-Grenoble, 38042 Grenoble, France.}	

\date{\today}

\begin{abstract}

Using typical experimental techniques it is difficult to separate the effects of carrier density from disorder on a two-dimensional superconducting transition.  To address this problem, we have produced graphene sheets decorated with a non-percolating network of nanoscale tin clusters.  These metal clusters both efficiently dope the graphene substrate and induce long-range superconducting correlations.  This allows us to study the superconducting transition at fixed disorder and variable carrier concentration via the field effect.  We find that despite structural inhomogeneity on mesoscopic length scales (10-100 nm), this material behaves electronically as a homogenous dirty superconductor.  Our simple self-assembly method establishes graphene as an ideal tunable substrate for studying induced two-dimensional electronic systems at fixed disorder and our technique can readily be extended to other order parameters such as magnetism.

\end{abstract}


\maketitle

The superconducting transition in two-dimensions is of interest for both the fundamental understanding of electronic order in reduced dimensions and applications involving superconducting thin films.  An open question is how the transition behaves as a the density of carriers mediating the superconductivity is varied.  In particular, the strength of disorder appears to play a fundamental role separating qualitatively different behavior\cite{Kapitulnik+Phase+Diagram}.  In two-dimensions, the electric field-effect provides the most versatile method for tuning the carrier density of a system at fixed disorder.  However, the field effect places more stringent limits on the dimensionality of the system since the film must be thinner than the Debye length, which is typically much smaller than the penetration depth that places the limit on two-dimensional superconductors.  Despite this limitation, the field effect has been used to tune the superconducting transition in specific materials such as thin films with anomalously low carrier density\cite{Goldman_FET_Bi} and interfacial states between complex oxides\cite{Triscone_FET}.  An alternative approach is to couple superconducting correlations directly into a truly two-dimensional electronic system, graphene.

The bipolar two-dimensional electron gas (2DEG) present in graphene\cite{Geim_Rev} is markedly different from the buried 2DEGs found at oxide interfaces or in GaAs heterostructures in that it is `open' to the environment with a stable and inert surface.  Using standard 2DEGs it is only feasible to capacitively couple the electron gas to materials deposited on their surface\cite{Clarke_DPT,Wagenblast+Dissipation+Theory}, whereas graphene is expected to allow direct coupling and thus offer access to different regions of phase space\cite{Kim_QPT_graphene}.  We sought to determine whether bulk materials deposited directly onto the graphene surface, such as metal clusters, can act as dopants and efficiently couple through the electron gas, whose carrier density and type can be tuned by an applied gate voltage.  The low carrier density in graphene, relative to bulk values, and weak intrinsic interactions such as spin-orbit coupling, should limit backaction of the electron gas on dopant properties.  Thus, exposed graphene sheets could provide a near-ideal substrate for the manipulation and general study of proximity-induced electronic phases.

Graphene has been shown to effectively carry proximity-induced Josephson currents injected from contacting electrodes\cite{Pablo_JJ,Andrei_JJ}.  However, a finite coherence length limits the length of such junctions to approximately one micron, reducing the physics to one dimension.  To maintain coherence over longer distances in two dimensions while retaining the unique properties of the graphene sheet we employ a geometry (Fig. 1a) where a large array of nanoscale dopant islands is placed in a non-percolating network on top of the graphene sheet\cite{Feigelman_graphene}.

We avoid complicated lithographic patterning and exploit the poor wettability of graphite to simply and reliably produce an array of submicron islands.  Low melting point metals such as the elemental superconductor Sn readily form self-assembled islands when deposited on pristine graphene at room temperature (Fig. 1b) similar to previous results on graphite\cite{Metois_Pb}.  Analysis of scanning electron micrographs and atomic force micrographs indicates that 10 nm of nominal deposition thickness typically results in islands with $80 \pm 5$ nm diameter  and $25 \pm 10$ nm gaps between them (Fig 1b).  In general, many different materials with different electronic order parameters can be deposited via this process by controlling the graphene substrate temperature during deposition\cite{Zayed_recrystal+In}, and other deposition methods such as chemical functionalization and wet self-assembly could be used as well.

Samples were prepared by exfoliating Kish graphite\cite{Geim_Rev} onto degenerately doped ($\rho < 0.005$ m$\Omega$-cm) silicon wafers coated with 285 nm of thermal oxide.  Single-layer graphene flakes were identified by optical contrast and confirmed via analysis with a micro-Raman spectrometer\cite{Ferrari_Raman}.  Four-probe contacts were defined via electron beam lithography and a Pd/Au 10/50 nm bilayer was evaporated as metal electrodes (Fig. 1, c). To produce the island network, Sn (99.999\% purity) was evaporated using an electron gun in high vacuum ($10^{-7}$ torr) onto graphene substrates at room temperature. The samples were thermally anchored to the cold stage of He-3 or Dilution cryostats and connected to highly filtered lines. Linear response and differential conductance were measured with standard low frequency lock-in techniques using low excitation currents in the range 10-100 nA. Figure 1c displays the room-temperature field-effect characteristics of a device before and after Sn is deposited.  Although 40\% of the graphene surface is coated by Sn islands after the deposition, many of the original electronic properties of graphene remain intact, including bipolar transport and field-effect mobilities $\mu > 1000$ $\mbox{cm}^2/(\mbox{V} \cdot \mbox{s})$.  The three main effects of Sn deposition are a rigid shift in the charge neutrality point (Dirac point $V_D$) to more negative voltages, a factor of five decrease in mobility, and a pronounced asymmetry between electron and hole transport.  Note that the maximum resistance at charge neutrality remains unchanged, indicating that the high coverage of low resistance Sn islands does not directly shunt the current in the graphene sheet. 

\begin{figure}
\vspace{-0.5cm}
\includegraphics[width=0.45\textwidth]{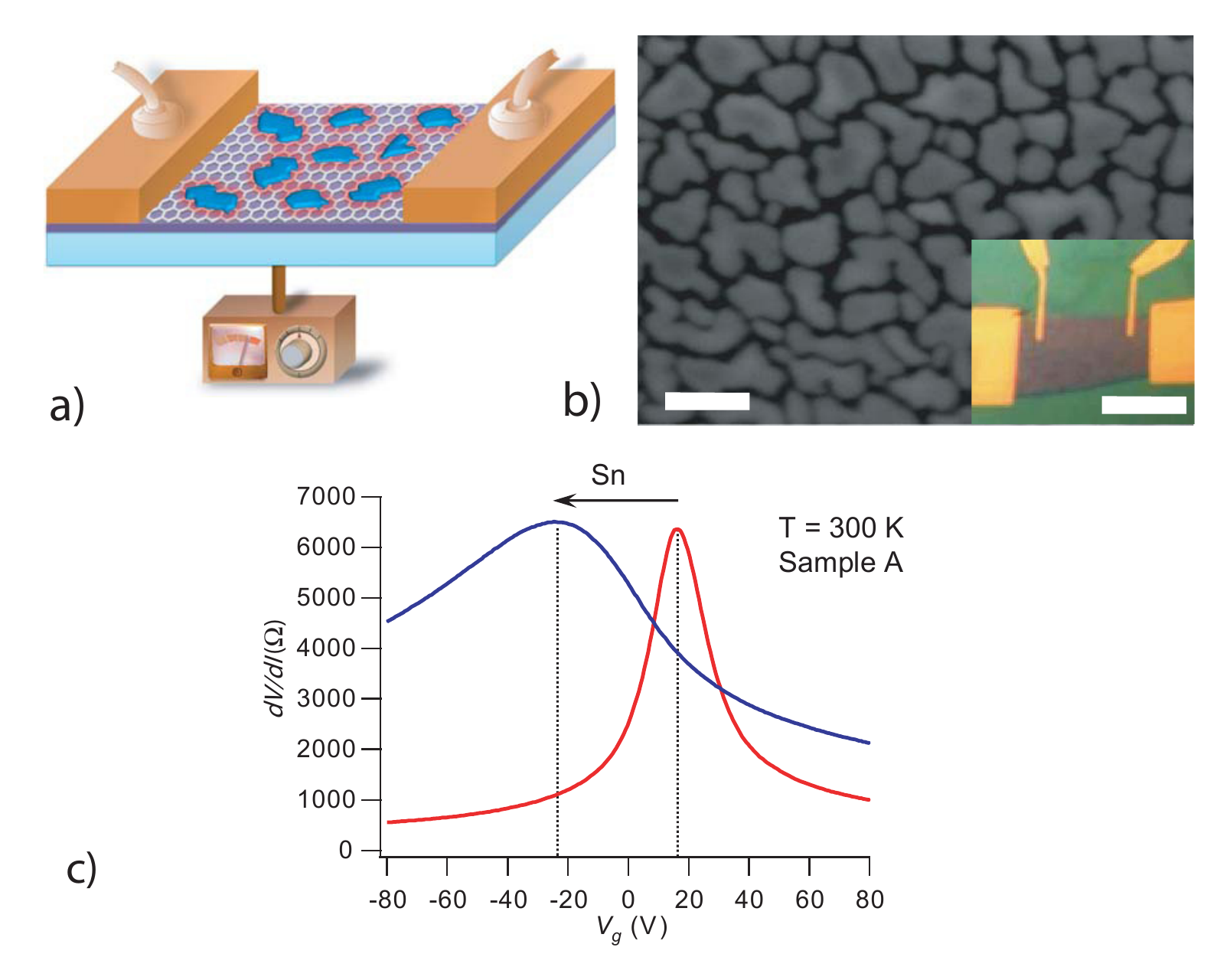}
\caption{\label{Fig. 1} (Color online) a) Schematic of device configuration and measurement setup.  Blue islands correspond to Sn clusters. b) Scanning electron micrograph of Sn island morphology on the graphene sheet (Scale bar = 100 nm.  Inset:  optical image of a typical device showing the four probe configuration  (Scale bar 10 microns) c) Four-terminal sheet resistance as a function of gate voltage for Sample A before (red online) and after (blue online) Sn deposition.  The dotted lines indicate the charge neutral point and the arrow indicates the shift after Sn deposition.}
\label{setup}
\end{figure}

All three of these effects are well described by inhomogeneous doping due to charge transfer from the metal islands to the graphene sheet\cite{Avouris_assymetry}.  From the shift in the charge neutrality point ($V_D$) on three separate samples and the known gate capacitance ($C_g$ = 115 aF / $\mu \mbox{m}^2$) we can calculate the charge induced in the graphene sheet by the Sn, $n_{ind} = C_g \Delta V_D$.  Normalizing by the observed Sn coverage, we infer that Sn transfers  $9 \pm 2 \times 10^{12}$ $\mbox{cm}^{-2}$ electrons to the graphene underneath it.  This is expected from the difference in work functions between the two materials ($\Phi_G = 4.5$ eV,  $\Phi_{Sn} = 4.42$ eV)\cite{Kelly_doping+contacts} and in agreement with recent experiments performed using other metals\cite{Kern_Contact_Effects}.  This induced charge reduces the mobility of both types of carriers via charged impurity scattering\cite{Fuhrer_Impurities} while the asymmetry in transport occurs because holes experience the pinned Fermi level under the Sn islands as a potential barrier, while electrons experience a potential well\cite{Avouris_assymetry,Goldhaber-Gordon_contacts}.

More interesting than the influence of the Sn islands on the normal state properties of graphene is the effect the superconducting correlations in the Sn have on transport via the proximity effect.  Figure 2 shows the sheet resistance versus temperature for gate voltages on both the hole (Fig. 2a) and electron (Fig. 2b) sides of the charge neutrality point.  Each curve exhibits two distinct features, a high temperature partial drop in resistance that occurs at $\sim 3.5$ K independent of gate voltage, and a broad transition between 3 K and 1 K to a state of zero resistance that is strongly dependent on the gate voltage.

\begin{figure}[bottom]
\includegraphics[width=0.40\textwidth]{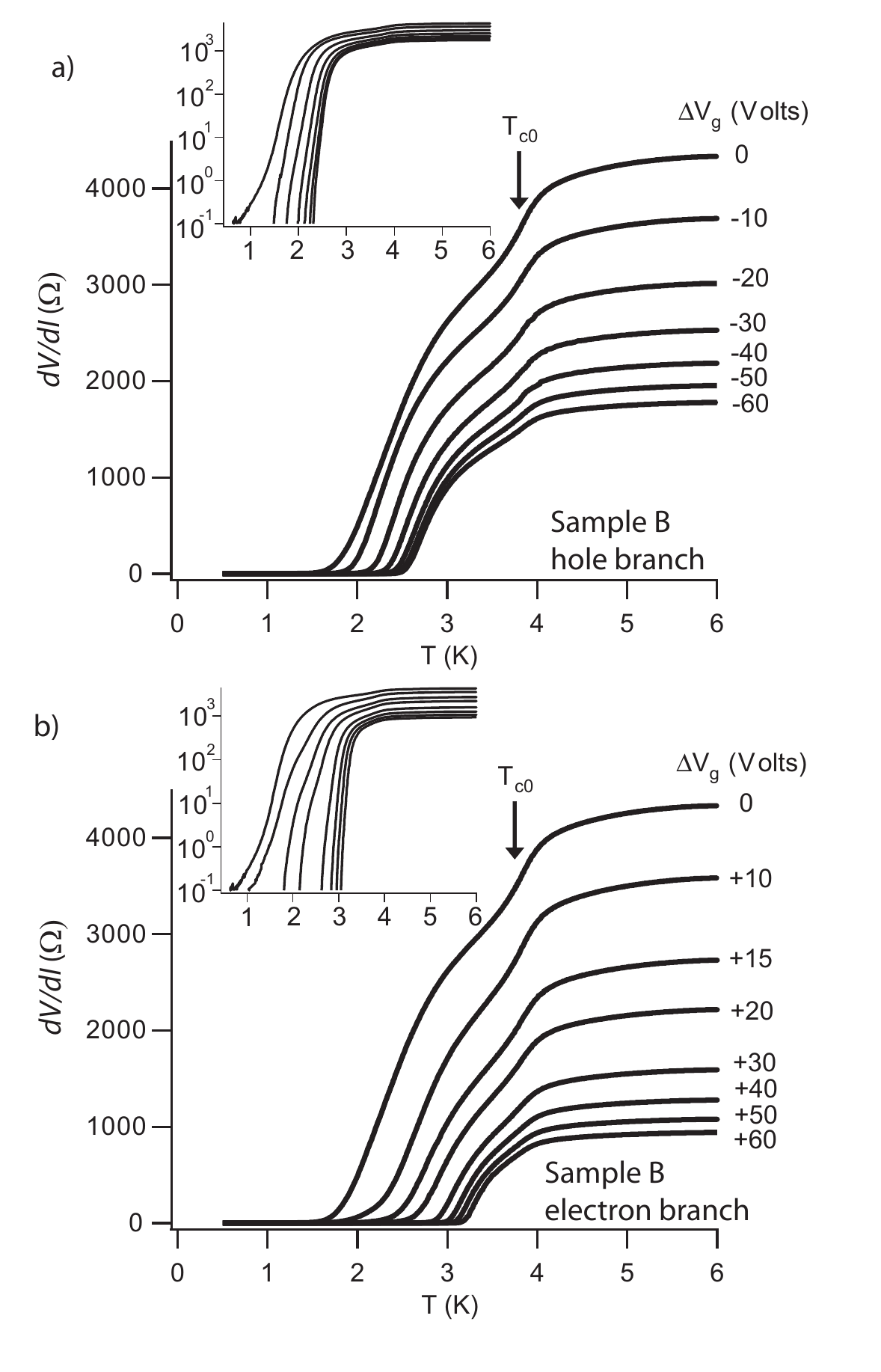}
\caption{\label{Fig. 2} Sheet resistance versus temperature for various gate voltages, $V_g$, referenced to the charge neutrality point $V_D = +40$ V for this device.  In a)  $\Delta V_g = V_g-V_D < 0$ corresponds to hole transport, whereas,  $\Delta V_g > 0$ in b) corresponds to electron transport through the graphene sheet.  The arrow labeled $T_{c0}$  indicates the first partial resistance drop corresponding to the mean-field pairing transition of the Sn islands. Inset: Same data on a log scale.}
\label{transition}
\end{figure}

The first partial resistance drop (arrows in Fig. 2a,b) is due to condensation of Cooper pairs in the Sn islands ($T^{Bulk}_c=3.72$ K). Analysis of the drop shows that it can be fit by $\Delta \sigma(T) \propto \ln (T/T_{c0})^{-1}$ (Fig. 3a) typical of Aslamazov-Larkin fluctuation-enhanced conductivity\cite{Aslamasoz+Larkin} in two dimensions.  Fitting each curve from $3.8-4.5$ K we extract a mean-field pairing temperature ($T_{c0}$) of $3.54\pm0.02$ K independent of gate voltage (solid squares in Fig. 3c).  Note that the amplitude of this drop is not directly proportional to the Sn coverage, indicating that the islands do not act as simple superconducting shunts.  In two-dimensional superconducting systems, it is well known that although the amplitude of the superconducting wave function is well defined below the pairing temperature $T_{c0}$, thermally induced phase fluctuations (vortices) destroy global phase coherence and produce dissipation due to a finite flux flow resistance\cite{Berezinskii2,Kosterlitz+Thouless}.  However, below the critical Berezinskii-Kosterlitz-Thouless unbinding temperature, $T_{BKT}$, the attractive interaction between vortices with opposite orientation causes them to form bound pairs allowing a finite supercurrent to flow.

The vortex-unbinding temperature can be identified from the universal form of the flux flow resistance\cite{Minnhagen_Rev} above the transition $R_{\Box} (T) \propto \exp[  b (T- T_{BKT})^{-1/2} ]$, where $b$ is a constant of order unity governing the vortex-antivortex interaction strength and $T_{BKT}$ is the vortex unbinding temperature.  To extract this form we plot $\left( d \ln(R_{\Box})/dT \right)^{-2/3}$ vs. T, which produces a straight line with $T_{BKT}$ given by the x-intercept for curves following the universal form.  In Fig. 3b the resulting fits are given showing $T_{BKT}$ extracted from the x-intercepts for three different gate voltages.  Through a large intermediate range the curves follow the universal form (straight line).  However, at low temperatures the curves level off due to finite-size effects, which cut off the attractive vortex-antivortex interaction\cite{Minnhagen_Rev}.  This departure from the universal form is particularly evident near the charge neutrality point (Fig. 2a and b), possibly indicating proximity to a superconductor-normal quantum critical point at the lowest charge densities\cite{Kivelson+SNT}.   Figure 3c summarizes the resulting $T_{BKT}$ (open circles) extracted from the resistance versus temperature curves at each gate voltage.

To analyze the gate voltage dependence of the vortex unbinding transition, we consider the sheet as a dirty two-dimensional superconductor where the gate voltage allows us to tune the normal state resistance.  This is justified since the length scales of disorder are much smaller than the superconducting coherence length, ie. $r \sim d \sim \ell_{mfp} < \xi_0$, where $r$ is the size of the islands, $d$ is the distance between islands, $ \ell_{mfp} \sim 20-30$ nm is the mean free path extracted from field effect measurements at 6 K and the superconducting coherence length Sn $( \xi^{Sn}_0 \sim 300$ nm).  For a dirty 2D superconductor, one can use the jump in superfluid stiffness at the vortex unbinding transition to relate $T_{BKT}$ to the normal state resistance of the film\cite{Beasley_BKT,Goldman_RG1},

\begin{equation}
\frac{ T_{c0} }{ T_{BKT} } \left\{ \frac{ \Delta(T_{BKT}) }{\Delta(0)} \tanh \left[ \frac{ \Delta(T_{BKT}) }{2k_b T_{BKT} } \right] \right\} = \frac{ \epsilon_v R_N }{ R_0 }
\end{equation}		 

where  $\Delta(T)$ is the superconducting energy gap, $R_N$ is the normal state sheet resistance, $R_0=\frac{2.18 \hbar}{e^2}\approx 8.96 k\Omega$, and $\epsilon_v$  is an effective dielectric constant that describes the material dependent screening of the attractive vortex-antivortex interaction\cite{Goldman_RG1}.  Using the weak-coupling BCS limit for the superconducting gap and the sheet resistance measured at 6 K, we fit the $T_{BKT}$ extracted above using $\epsilon_v$ as the only adjustable parameter and find $\epsilon_v=2.40\pm0.05$ (solid curve in Fig.3), which is twice the value seen in related systems\cite{Goldman_RG1}.  This implies that the vortices and antivortices are relatively weakly bound in our system.

\begin{figure}
\includegraphics[width=0.49\textwidth]{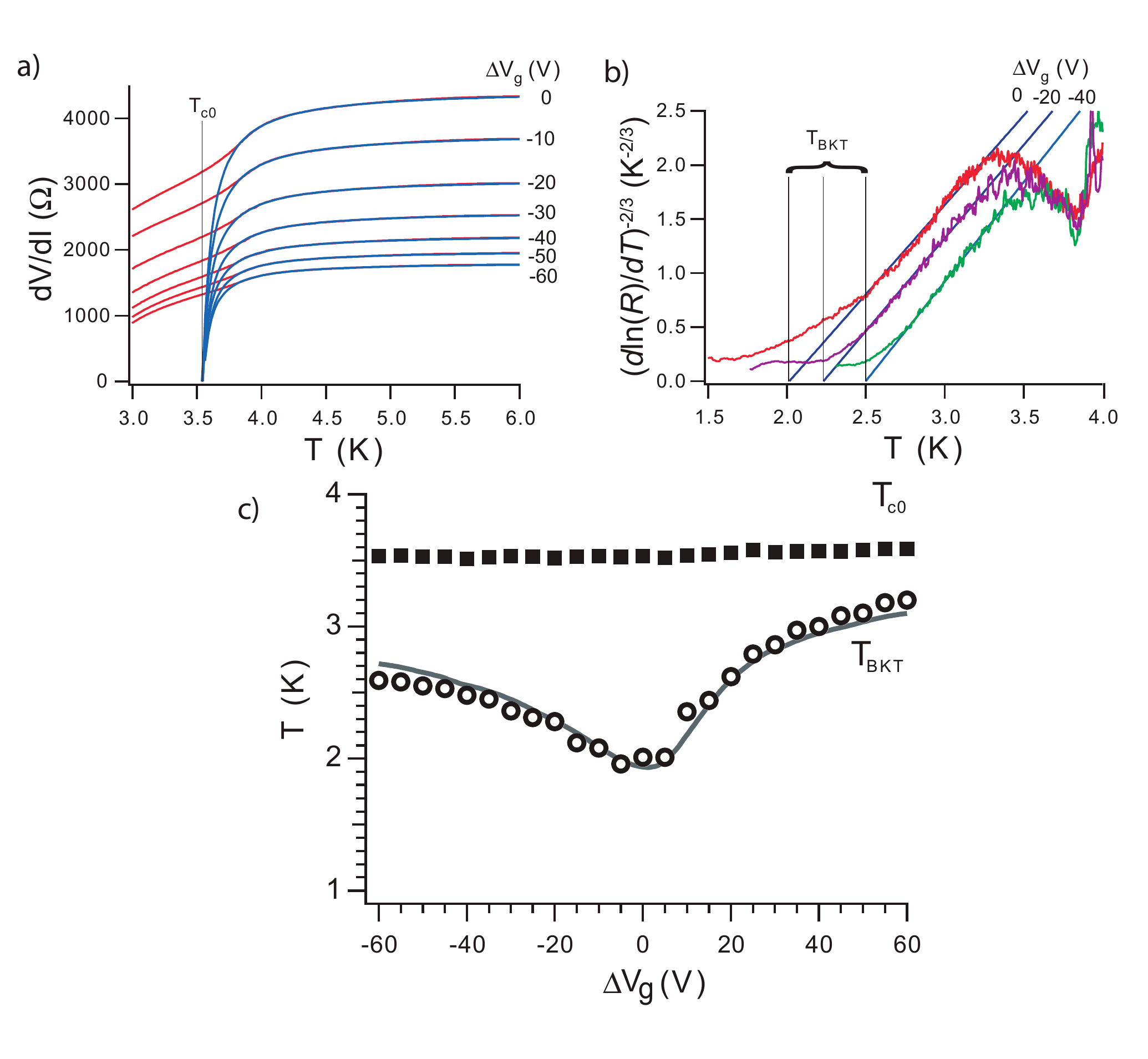}
\caption{\label{Fig. 3} (Color online) a) Fits of the sheet resistance versus temperature to fluctuation-enhanced conductivity of the Aslamazov-Larkin form.  b) Rescaling of the sheet resistance versus temperature to the BKT form to extract the vortex unbinding temperature $T_{BKT}$ c) The mean-field pairing temperature, $T_{c0}$ (black squares) and vortex-unbinding temperature, $T_{BKT}$ (open circles) as a function of gate voltage.  The solid line is a fit of $T_{BKT}$ using equation 1 and the measured normal state properties of the device.  (see text).}
\label{fitsummary}
\end{figure}

At temperatures below $T_{BKT}$, vortices and antivortices form bound pairs and a finite critical current develops which saturates to a gate-voltage dependent value for $T \ll T_{BKT}$.  Current-voltage characteristics at 100 mK for different applied gate voltages are shown in Figure 4.  The gate-tunable critical current is qualitatively similar to isolated graphene Josephson junctions\cite{Pablo_JJ,Andrei_JJ} with the exception that, in our devices, critical current densities ($I_c/width \sim 1$ A/m) comparable to submicron graphene Josephson junctions are maintained over distances of tens of microns, demonstrating the fully two-dimensional phase coherence in this system.

We have demonstrated a simple method to produce a two-dimensional superconductor on a graphene substrate and tune the transition via an electrostatic gate.  This allowed us to systematically tune the carrier density at fixed disorder.  Although structurally inhomogenous, this material behaved electronically as a weakly disordered two-dimensional superconductor.  While we have probed the properties of this system using electron transport, the readily accessible interface allows application of a myriad of local characterization techniques such as scanning probe microscopy, optical spectroscopy, etc.  We expect arranging the islands into regular arrays or superlattices should lead to interesting frustration effects as a function of applied magnetic and electric fields\cite{Jayaprakash_JJ_in_B,Louie_superlattices}.  This tunable superconducting material may find applications in bolometers for sensing applications or as an element in circuits for quantum information processing.  The ease of fabrication and considerable versatility of deposition materials make graphene an attractive platform for investigating other electronic orders such as magnetism in two dimensions.

\begin{figure}
\vspace{-1cm}
\includegraphics[width=0.45\textwidth]{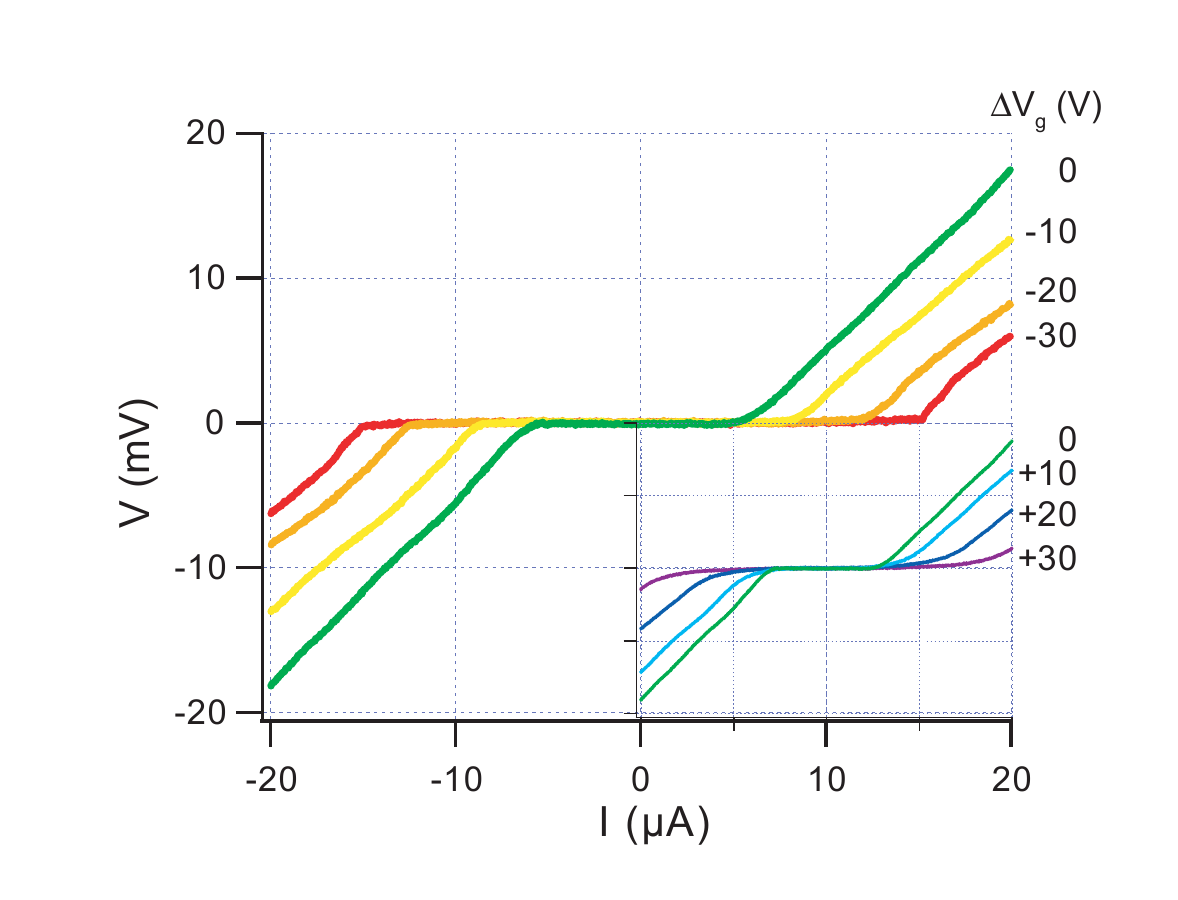}
\caption{\label{Fig. 4} (Color online) Current-voltage curves taken at zero magnetic field and 100 mK, corresponding to hole transport in the graphene sheet for gate voltages, $\Delta V_g$ relative to the Dirac point $V_D=+12$ V for this sample. Inset: similiar curves for electron transport. }
\label{groundstate}
\end{figure}

\begin{acknowledgments}
B.K., \c{C}.G., and A.Z. were supported by the Director, Office of Energy Research, Office of Basic Energy Sciences, Materials Sciences, and Engineering Division, of the U.S. Department of Energy under contract DE-AC02-05CH11231, through the sp2-bonded nanostructures program.  V.B. acknowledges support from the Miller Institute for Basic Research in Science, CNRS/MPPU and ANR-JC/NEMESIS. We thank M. Feigelman, M. Skvortsov, J. Moore, G. Deutscher and P. Ghaemi for helpful discussions.
\end{acknowledgments}



\begin{thebibliography}{27}
\expandafter\ifx\csname natexlab\endcsname\relax\def\natexlab#1{#1}\fi
\expandafter\ifx\csname bibnamefont\endcsname\relax
  \def\bibnamefont#1{#1}\fi
\expandafter\ifx\csname bibfnamefont\endcsname\relax
  \def\bibfnamefont#1{#1}\fi
\expandafter\ifx\csname citenamefont\endcsname\relax
  \def\citenamefont#1{#1}\fi
\expandafter\ifx\csname url\endcsname\relax
  \def\url#1{\texttt{#1}}\fi
\expandafter\ifx\csname urlprefix\endcsname\relax\def\urlprefix{URL }\fi
\providecommand{\bibinfo}[2]{#2}
\providecommand{\eprint}[2][]{\url{#2}}

\bibitem[{\citenamefont{Steiner et~al.}(2008)\citenamefont{Steiner, Breznay,
  and Aharon}}]{Kapitulnik+Phase+Diagram}
\bibinfo{author}{\bibfnamefont{M.~A.} \bibnamefont{Steiner}},
  \bibinfo{author}{\bibfnamefont{N.~P.} \bibnamefont{Breznay}},
  \bibnamefont{and} \bibinfo{author}{\bibfnamefont{K.}~\bibnamefont{Aharon}},
  \bibinfo{journal}{Phys. Rev. B} \textbf{\bibinfo{volume}{77}},
  \bibinfo{pages}{212501} (\bibinfo{year}{2008}).

\bibitem[{\citenamefont{Parendo et~al.}(2005)\citenamefont{Parendo, Sarwa, Tan,
  Bhattacharya, Eblen-Zayas, Staley, and Goldman}}]{Goldman_FET_Bi}
\bibinfo{author}{\bibfnamefont{K.~A.} \bibnamefont{Parendo}},
  \bibinfo{author}{\bibfnamefont{K.~H.} \bibnamefont{Sarwa}},
  \bibinfo{author}{\bibfnamefont{B.}~\bibnamefont{Tan}},
  \bibinfo{author}{\bibfnamefont{A.}~\bibnamefont{Bhattacharya}},
  \bibinfo{author}{\bibfnamefont{M.}~\bibnamefont{Eblen-Zayas}},
  \bibinfo{author}{\bibfnamefont{N.~E.} \bibnamefont{Staley}},
  \bibnamefont{and} \bibinfo{author}{\bibfnamefont{A.~M.}
  \bibnamefont{Goldman}}, \bibinfo{journal}{Phys. Rev. Lett.}
  \textbf{\bibinfo{volume}{94}}, \bibinfo{pages}{197004}
  (\bibinfo{year}{2005}).

\bibitem[{\citenamefont{Caviglia et~al.}(2008)\citenamefont{Caviglia, Gariglio,
  Reyren, Jaccard, Schneider, Gabay, Thiel, Hammerl, Mannhart, and
  Triscone}}]{Triscone_FET}
\bibinfo{author}{\bibfnamefont{A.~D.} \bibnamefont{Caviglia}},
  \bibinfo{author}{\bibfnamefont{S.}~\bibnamefont{Gariglio}},
  \bibinfo{author}{\bibfnamefont{N.}~\bibnamefont{Reyren}},
  \bibinfo{author}{\bibfnamefont{D.}~\bibnamefont{Jaccard}},
  \bibinfo{author}{\bibfnamefont{T.}~\bibnamefont{Schneider}},
  \bibinfo{author}{\bibfnamefont{M.}~\bibnamefont{Gabay}},
  \bibinfo{author}{\bibfnamefont{S.}~\bibnamefont{Thiel}},
  \bibinfo{author}{\bibfnamefont{G.}~\bibnamefont{Hammerl}},
  \bibinfo{author}{\bibfnamefont{J.}~\bibnamefont{Mannhart}}, \bibnamefont{and}
  \bibinfo{author}{\bibfnamefont{J.~M.} \bibnamefont{Triscone}},
  \bibinfo{journal}{Nature} \textbf{\bibinfo{volume}{456}},
  \bibinfo{pages}{624} (\bibinfo{year}{2008}).

\bibitem[{\citenamefont{Geim and Novoselov}(2007)}]{Geim_Rev}
\bibinfo{author}{\bibfnamefont{A.~K.} \bibnamefont{Geim}} \bibnamefont{and}
  \bibinfo{author}{\bibfnamefont{K.~S.} \bibnamefont{Novoselov}},
  \bibinfo{journal}{Nat. Mater.} \textbf{\bibinfo{volume}{6}},
  \bibinfo{pages}{183} (\bibinfo{year}{2007}).

\bibitem[{\citenamefont{Rimberg et~al.}(1997)\citenamefont{Rimberg, Ho, Kurdak,
  Clarke, Campman, and Gossard}}]{Clarke_DPT}
\bibinfo{author}{\bibfnamefont{A.~J.} \bibnamefont{Rimberg}},
  \bibinfo{author}{\bibfnamefont{T.~R.} \bibnamefont{Ho}},
  \bibinfo{author}{\bibfnamefont{C.}~\bibnamefont{Kurdak}},
  \bibinfo{author}{\bibfnamefont{J.}~\bibnamefont{Clarke}},
  \bibinfo{author}{\bibfnamefont{K.~L.} \bibnamefont{Campman}},
  \bibnamefont{and} \bibinfo{author}{\bibfnamefont{A.~C.}
  \bibnamefont{Gossard}}, \bibinfo{journal}{Phys. Rev. Lett.}
  \textbf{\bibinfo{volume}{78}}, \bibinfo{pages}{2632} (\bibinfo{year}{1997}).

\bibitem[{\citenamefont{Wagenblast et~al.}(1997)\citenamefont{Wagenblast, van
  Otterlo, Sch\"on, and Zim\'anyi}}]{Wagenblast+Dissipation+Theory}
\bibinfo{author}{\bibfnamefont{K.-H.} \bibnamefont{Wagenblast}},
  \bibinfo{author}{\bibfnamefont{A.}~\bibnamefont{van Otterlo}},
  \bibinfo{author}{\bibfnamefont{G.}~\bibnamefont{Sch\"on}}, \bibnamefont{and}
  \bibinfo{author}{\bibfnamefont{G.~T.} \bibnamefont{Zim\'anyi}},
  \bibinfo{journal}{Phys. Rev. Lett.} \textbf{\bibinfo{volume}{79}},
  \bibinfo{pages}{2730} (\bibinfo{year}{1997}).

\bibitem[{\citenamefont{Takei and Kim}(2008)}]{Kim_QPT_graphene}
\bibinfo{author}{\bibfnamefont{S.}~\bibnamefont{Takei}} \bibnamefont{and}
  \bibinfo{author}{\bibfnamefont{Y.~B.} \bibnamefont{Kim}},
  \bibinfo{journal}{Phys. Rev. B} \textbf{\bibinfo{volume}{78}},
  \bibinfo{pages}{165401} (\bibinfo{year}{2008}).

\bibitem[{\citenamefont{Heersche et~al.}(2007)\citenamefont{Heersche,
  Jarillo-Herrero, Oostinga, Vandersypen, and Morpurgo}}]{Pablo_JJ}
\bibinfo{author}{\bibfnamefont{H.~B.} \bibnamefont{Heersche}},
  \bibinfo{author}{\bibfnamefont{P.}~\bibnamefont{Jarillo-Herrero}},
  \bibinfo{author}{\bibfnamefont{J.~B.} \bibnamefont{Oostinga}},
  \bibinfo{author}{\bibfnamefont{L.~M.~K.} \bibnamefont{Vandersypen}},
  \bibnamefont{and} \bibinfo{author}{\bibfnamefont{A.~F.}
  \bibnamefont{Morpurgo}}, \bibinfo{journal}{Nature}
  \textbf{\bibinfo{volume}{446}}, \bibinfo{pages}{56} (\bibinfo{year}{2007}).

\bibitem[{\citenamefont{Du et~al.}(2008)\citenamefont{Du, Skachko, and
  Andrei}}]{Andrei_JJ}
\bibinfo{author}{\bibfnamefont{X.}~\bibnamefont{Du}},
  \bibinfo{author}{\bibfnamefont{I.}~\bibnamefont{Skachko}}, \bibnamefont{and}
  \bibinfo{author}{\bibfnamefont{E.~Y.} \bibnamefont{Andrei}},
  \bibinfo{journal}{Physical Review B} \textbf{\bibinfo{volume}{77}},
  \bibinfo{pages}{184507} (\bibinfo{year}{2008}).

\bibitem[{\citenamefont{Feigel'man et~al.}(2008)\citenamefont{Feigel'man,
  Skvortsov, and Tikhonov}}]{Feigelman_graphene}
\bibinfo{author}{\bibfnamefont{M.~V.} \bibnamefont{Feigel'man}},
  \bibinfo{author}{\bibfnamefont{M.~A.} \bibnamefont{Skvortsov}},
  \bibnamefont{and} \bibinfo{author}{\bibfnamefont{K.~S.}
  \bibnamefont{Tikhonov}}, \bibinfo{journal}{JETP Lett.}
  \textbf{\bibinfo{volume}{88}}, \bibinfo{pages}{862} (\bibinfo{year}{2008}).

\bibitem[{\citenamefont{Heyraud and M\'etois}(1983)}]{Metois_Pb}
\bibinfo{author}{\bibfnamefont{J.~C.} \bibnamefont{Heyraud}} \bibnamefont{and}
  \bibinfo{author}{\bibfnamefont{J.~J.} \bibnamefont{M\'etois}},
  \bibinfo{journal}{Surface Science} \textbf{\bibinfo{volume}{128}},
  \bibinfo{pages}{334} (\bibinfo{year}{1983}).

\bibitem[{\citenamefont{Zayed and Elsayed-Ali}(2005)}]{Zayed_recrystal+In}
\bibinfo{author}{\bibfnamefont{M.~K.} \bibnamefont{Zayed}} \bibnamefont{and}
  \bibinfo{author}{\bibfnamefont{H.~E.} \bibnamefont{Elsayed-Ali}},
  \bibinfo{journal}{Thin Solid Films} \textbf{\bibinfo{volume}{489}},
  \bibinfo{pages}{42} (\bibinfo{year}{2005}).

\bibitem[{\citenamefont{Ferrari et~al.}(2006)\citenamefont{Ferrari, Meyer,
  Scardaci, Casiraghi, Lazzeri, Mauri, Piscanec, Jiang, Novoselov, Roth
  et~al.}}]{Ferrari_Raman}
\bibinfo{author}{\bibfnamefont{A.~C.} \bibnamefont{Ferrari}},
  \bibinfo{author}{\bibfnamefont{J.~C.} \bibnamefont{Meyer}},
  \bibinfo{author}{\bibfnamefont{V.}~\bibnamefont{Scardaci}},
  \bibinfo{author}{\bibfnamefont{C.}~\bibnamefont{Casiraghi}},
  \bibinfo{author}{\bibfnamefont{M.}~\bibnamefont{Lazzeri}},
  \bibinfo{author}{\bibfnamefont{F.}~\bibnamefont{Mauri}},
  \bibinfo{author}{\bibfnamefont{S.}~\bibnamefont{Piscanec}},
  \bibinfo{author}{\bibfnamefont{D.}~\bibnamefont{Jiang}},
  \bibinfo{author}{\bibfnamefont{K.~S.} \bibnamefont{Novoselov}},
  \bibinfo{author}{\bibfnamefont{S.}~\bibnamefont{Roth}}, \bibnamefont{et~al.},
  \bibinfo{journal}{Phys. Rev. Lett.} \textbf{\bibinfo{volume}{97}},
  \bibinfo{pages}{187401} (\bibinfo{year}{2006}).

\bibitem[{\citenamefont{Farmer et~al.}(2009)\citenamefont{Farmer,
  Golizadeh-Mojarad, Perebeinos, Lin, Tulevski, Tsang, and
  Avouris}}]{Avouris_assymetry}
\bibinfo{author}{\bibfnamefont{D.~B.} \bibnamefont{Farmer}},
  \bibinfo{author}{\bibfnamefont{R.}~\bibnamefont{Golizadeh-Mojarad}},
  \bibinfo{author}{\bibfnamefont{V.}~\bibnamefont{Perebeinos}},
  \bibinfo{author}{\bibfnamefont{Y.~M.} \bibnamefont{Lin}},
  \bibinfo{author}{\bibfnamefont{G.~S.} \bibnamefont{Tulevski}},
  \bibinfo{author}{\bibfnamefont{J.~C.} \bibnamefont{Tsang}}, \bibnamefont{and}
  \bibinfo{author}{\bibfnamefont{P.}~\bibnamefont{Avouris}},
  \bibinfo{journal}{Nano Lett.} \textbf{\bibinfo{volume}{9}},
  \bibinfo{pages}{388} (\bibinfo{year}{2009}).

\bibitem[{\citenamefont{Giovannetti et~al.}(2008)\citenamefont{Giovannetti,
  Khomyakov, Brocks, Karpan, van~den Brink, and Kelly}}]{Kelly_doping+contacts}
\bibinfo{author}{\bibfnamefont{G.}~\bibnamefont{Giovannetti}},
  \bibinfo{author}{\bibfnamefont{P.~A.} \bibnamefont{Khomyakov}},
  \bibinfo{author}{\bibfnamefont{G.}~\bibnamefont{Brocks}},
  \bibinfo{author}{\bibfnamefont{V.~M.} \bibnamefont{Karpan}},
  \bibinfo{author}{\bibfnamefont{J.}~\bibnamefont{van~den Brink}},
  \bibnamefont{and} \bibinfo{author}{\bibfnamefont{P.~J.} \bibnamefont{Kelly}},
  \bibinfo{journal}{Phys. Rev. Lett.} \textbf{\bibinfo{volume}{101}},
  \bibinfo{pages}{026803} (\bibinfo{year}{2008}).

\bibitem[{\citenamefont{Lee et~al.}(2008)\citenamefont{Lee, Balasubramanian,
  Weitz, Burghard, and Kern}}]{Kern_Contact_Effects}
\bibinfo{author}{\bibfnamefont{E.~J.~H.} \bibnamefont{Lee}},
  \bibinfo{author}{\bibfnamefont{K.}~\bibnamefont{Balasubramanian}},
  \bibinfo{author}{\bibfnamefont{R.~T.} \bibnamefont{Weitz}},
  \bibinfo{author}{\bibfnamefont{M.}~\bibnamefont{Burghard}}, \bibnamefont{and}
  \bibinfo{author}{\bibfnamefont{K.}~\bibnamefont{Kern}},
  \bibinfo{journal}{Nat. Nanotechnol.} \textbf{\bibinfo{volume}{3}},
  \bibinfo{pages}{486} (\bibinfo{year}{2008}).

\bibitem[{\citenamefont{Chen et~al.}(2008)\citenamefont{Chen, Jang, Adam,
  Fuhrer, Williams, and Ishigami}}]{Fuhrer_Impurities}
\bibinfo{author}{\bibfnamefont{J.~H.} \bibnamefont{Chen}},
  \bibinfo{author}{\bibfnamefont{C.}~\bibnamefont{Jang}},
  \bibinfo{author}{\bibfnamefont{S.}~\bibnamefont{Adam}},
  \bibinfo{author}{\bibfnamefont{M.~S.} \bibnamefont{Fuhrer}},
  \bibinfo{author}{\bibfnamefont{E.~D.} \bibnamefont{Williams}},
  \bibnamefont{and} \bibinfo{author}{\bibfnamefont{M.}~\bibnamefont{Ishigami}},
  \bibinfo{journal}{Nat. Phys.} \textbf{\bibinfo{volume}{4}},
  \bibinfo{pages}{377} (\bibinfo{year}{2008}).

\bibitem[{\citenamefont{Huard et~al.}(2008)\citenamefont{Huard, Stander,
  Sulpizio, and Goldhaber-Gordon}}]{Goldhaber-Gordon_contacts}
\bibinfo{author}{\bibfnamefont{B.}~\bibnamefont{Huard}},
  \bibinfo{author}{\bibfnamefont{N.}~\bibnamefont{Stander}},
  \bibinfo{author}{\bibfnamefont{J.~A.} \bibnamefont{Sulpizio}},
  \bibnamefont{and}
  \bibinfo{author}{\bibfnamefont{D.}~\bibnamefont{Goldhaber-Gordon}},
  \bibinfo{journal}{Phys. Rev. B} \textbf{\bibinfo{volume}{78}},
  \bibinfo{pages}{121402(R)} (\bibinfo{year}{2008}).

\bibitem[{\citenamefont{Aslamazov and Larkin}(1968)}]{Aslamasoz+Larkin}
\bibinfo{author}{\bibfnamefont{L.}~\bibnamefont{Aslamazov}} \bibnamefont{and}
  \bibinfo{author}{\bibfnamefont{A.~I.} \bibnamefont{Larkin}},
  \bibinfo{journal}{Phys. Lett. A} \textbf{\bibinfo{volume}{A 26}},
  \bibinfo{pages}{238} (\bibinfo{year}{1968}).

\bibitem[{\citenamefont{Berezinskii}(1972)}]{Berezinskii2}
\bibinfo{author}{\bibfnamefont{V.}~\bibnamefont{Berezinskii}},
  \bibinfo{journal}{Sov. Phys. JETP} \textbf{\bibinfo{volume}{34}},
  \bibinfo{pages}{610} (\bibinfo{year}{1972}).

\bibitem[{\citenamefont{Kosterlitz and Thouless}(1973)}]{Kosterlitz+Thouless}
\bibinfo{author}{\bibfnamefont{J.~M.} \bibnamefont{Kosterlitz}}
  \bibnamefont{and} \bibinfo{author}{\bibfnamefont{D.~J.}
  \bibnamefont{Thouless}}, \bibinfo{journal}{J. Phys. C: Solid State Phys.}
  \textbf{\bibinfo{volume}{6}}, \bibinfo{pages}{1181} (\bibinfo{year}{1973}).

\bibitem[{\citenamefont{Minnhagen}(1987)}]{Minnhagen_Rev}
\bibinfo{author}{\bibfnamefont{P.}~\bibnamefont{Minnhagen}},
  \bibinfo{journal}{Rev. Mod. Phys.} \textbf{\bibinfo{volume}{59}},
  \bibinfo{pages}{1001} (\bibinfo{year}{1987}).

\bibitem[{\citenamefont{Spivak et~al.}(2008)\citenamefont{Spivak, Oreto, and
  Kivelson}}]{Kivelson+SNT}
\bibinfo{author}{\bibfnamefont{B.}~\bibnamefont{Spivak}},
  \bibinfo{author}{\bibfnamefont{P.}~\bibnamefont{Oreto}}, \bibnamefont{and}
  \bibinfo{author}{\bibfnamefont{S.~A.} \bibnamefont{Kivelson}},
  \bibinfo{journal}{Phys. Rev. B} \textbf{\bibinfo{volume}{77}},
  \bibinfo{pages}{214523} (\bibinfo{year}{2008}).

\bibitem[{\citenamefont{Beasley et~al.}(1979)\citenamefont{Beasley, Mooij, and
  Orlando}}]{Beasley_BKT}
\bibinfo{author}{\bibfnamefont{M.~R.} \bibnamefont{Beasley}},
  \bibinfo{author}{\bibfnamefont{J.~E.} \bibnamefont{Mooij}}, \bibnamefont{and}
  \bibinfo{author}{\bibfnamefont{T.~P.} \bibnamefont{Orlando}},
  \bibinfo{journal}{Phys. Rev. Lett.} \textbf{\bibinfo{volume}{42}},
  \bibinfo{pages}{1165} (\bibinfo{year}{1979}).

\bibitem[{\citenamefont{Epstein et~al.}(1982)\citenamefont{Epstein, Goldman,
  and Kadin}}]{Goldman_RG1}
\bibinfo{author}{\bibfnamefont{K.}~\bibnamefont{Epstein}},
  \bibinfo{author}{\bibfnamefont{A.~M.} \bibnamefont{Goldman}},
  \bibnamefont{and} \bibinfo{author}{\bibfnamefont{A.~M.} \bibnamefont{Kadin}},
  \bibinfo{journal}{Physical Review B} \textbf{\bibinfo{volume}{26}},
  \bibinfo{pages}{3950} (\bibinfo{year}{1982}).

\bibitem[{\citenamefont{Teitel and Jayaprakash}(1983)}]{Jayaprakash_JJ_in_B}
\bibinfo{author}{\bibfnamefont{S.}~\bibnamefont{Teitel}} \bibnamefont{and}
  \bibinfo{author}{\bibfnamefont{C.}~\bibnamefont{Jayaprakash}},
  \bibinfo{journal}{Phys. Rev. Lett.} \textbf{\bibinfo{volume}{51}},
  \bibinfo{pages}{1999} (\bibinfo{year}{1983}).

\bibitem[{\citenamefont{Park et~al.}(2008)\citenamefont{Park, Yang, Son, Cohen,
  and Louie}}]{Louie_superlattices}
\bibinfo{author}{\bibfnamefont{C.~H.} \bibnamefont{Park}},
  \bibinfo{author}{\bibfnamefont{L.}~\bibnamefont{Yang}},
  \bibinfo{author}{\bibfnamefont{Y.~W.} \bibnamefont{Son}},
  \bibinfo{author}{\bibfnamefont{M.~L.} \bibnamefont{Cohen}}, \bibnamefont{and}
  \bibinfo{author}{\bibfnamefont{S.~G.} \bibnamefont{Louie}},
  \bibinfo{journal}{Nat. Phys.} \textbf{\bibinfo{volume}{4}},
  \bibinfo{pages}{213} (\bibinfo{year}{2008}).

\end{thebibliography}
\end{document}